\documentclass[a4paper]{article}

\usepackage[T1]{fontenc} 
\usepackage{float}       
\usepackage{helvet}      
\usepackage{mathptmx}    
\usepackage{rsc}         
\usepackage{setspace}    
\AtBeginDocument{\doublespacing}
\newfloat{chart}{htbp}{loc}
\newfloat{graph}{htbp}{loh}
\newfloat{scheme}{htbp}{los}

\usepackage{graphicx}
\usepackage{amssymb}
\usepackage{amsmath}
\usepackage{fullpage}

\title{Equilibration of granular subsystems}

\author{%
  F. Lechenault\thanks{%
    Department of Physics, NC State University, Raleigh, NC, USA,%
  }%
  \and
  Karen E. Daniels\thanks{%
    Department of Physics, NC State University, Raleigh, NC, USA,%
    E-mail: \texttt{kdaniel@ncsu.edu}
  }
}

\date{17 December, 2009}

\begin{document}
\maketitle

\begin{abstract}
We experimentally investigate the steady states of two granular assemblies
differing in their material properties and allowed to exchange volume with each
other under external agitation in the vicinity of their jamming transition.  We
extract the statistics of various static and dynamic quantities, and uncover a
materials-independent relationship between the average packing fraction and
its fluctuations. This relationship defines an intensive parameter which
decouples from the volume statistics, and remarkably takes the same value in
both subsystems. We also observe that an effective diffusion coefficient also
takes the same value in each subsystem, even as the structural relaxation time
increases over several orders of magnitude. These observations provide strong
constraints on the eventual establishment of a granular equation of state. 
\end{abstract} 


\section{Introduction}

Statistical mechanics has proven to be one of the most powerful tools for
understanding transitions between states in physical systems. 
This predictive power has been successfully extended to gently perturbed
systems through linear response theory.\citep{Kubo-1966-FDT} Such an
approach emphasizes the significance of {\it intensive}
state variables, analogous to temperature or pressure in equilibrium systems, 
but potentially extends
their relevance to the wider spectrum of out-of-equilibrium systems.
However, this formalism is not valid for systems driven
far from equilibrium, even when in well defined steady states. It remains
an open question whether or not state variables can be defined in a generic way
for non-equilibrium steady states. The hope is that such state variables would
carry predictive power regarding the system's response to a change in the
driving parameters, or its behavior when put in contact with another system. 

A key situation in which such tools are expected to be relevant is that of
granular materials: systems composed of individual particles large enough to be
insensitive to thermal fluctuations. Such materials are ubiquitous in natural
and industrial contexts, and models of their dynamics would benefit from
an improved understanding of how to prepare and manipulate them in well-defined
macrostates determined by a small set of control parameters analogous to
thermal intensivities. Promising measures of the temperature-like
Edwards compactivity \citep{Edwards-1989-TP,Coniglio-2005-SMD} 
have been made in static granular systems \citep{Nowak-1998-DFV, 
Schroter-2005-SSV, McNamara-2009-MGE}, as well as 
Kubo-type fluctuation/response pairs \citep{Cugliandolo-1997-EFP}
leading to effective temperature 
measurements in dynamic granular systems \citep{DAnna-2001-JRG, 
Xu-2005-ETA, Potiguar-2006-ETJ, Wang-2008-PDE}. However, 
it remains an open question to what extent these temperature-like variables 
behave as true temperatures.

For granular materials under strong agitation, the interactions of individual
particles are dominated by binary collisions.
Such dynamics are well-described by a thermal-like kinetic theory in which the
velocity fluctuations play the role of temperature
\citep{Jenkins-1983-TRF, Lun-1984-KTG}. However, the denser states relevant to
many important granular flows are dominated by multi-body interactions and such
states remain difficult to model through kinetic theory approaches
\citep{moon2001velocity}.
The phenomenology of these dense states is strikingly reminiscent of supercooled
liquids
\citep{Liu-1998-NDJ,Glotzer-2000-TDF,Silbert-2005-THD,Abate-2006-AJA,
Keys-2007-MGD,Lechenault-2008-CSH}, where classical statistical
mechanics also loses validity due to a loss of ergodicity. Both the
thermodynamic glass transition and the jamming transition in dense granular
materials lack a fundamental understanding, and both have been hampered by the
difficulty of making a relevant quantitative description of the state of the
system. This universal phenomenology, embodied by a dramatic slowing down of the
dynamics \citep{Mayor-2004-JWP}, 
caging \citep{Weeks-2002-PCR, Dauchot-2005-DHC, Reis-2007-CDG}, 
and memory effects \citep{Josserand-2000-MEG,Berthier-2002-GAA, 
Berthier-2002-SCL, Friedmann-2003-GMI}, 
is now referred to as the {\it jamming} transition, and
is at the heart of a very active field of research which is the subject of 
this special issue.

\bigskip

Thermodynamics is founded on, among other things, a meaningful equilibration of
intensive
state variables between subsystems.  Thus, if a thermodynamic-like description
of granular materials is to be found, an informative first step would be to
search
for an analog of the zeroth law of thermodynamics. By putting two athermal
systems in contact and allowing them to reach a steady state, one can seek
relevant intensive parameters by discarding properties that fail to equilibrate.
We experimentally conduct such an investigation, putting two model granular
subsystems in contact.

In our experiments, we allow two adjacent monolayers of particles, differing in
material properties, to exchange volume with each other under external agitation
and an overall constant-volume constraint. We study the steady-state properties
of the two subsystems as a function of the overall average packing fraction in
the vicinity of
their jamming transition. Importantly, we uncover a materials-independent
relationship between the average packing fraction
and its fluctuations. This relationship defines an intensive parameter that
decouples from the volume statistics of the subsystems, and remarkably takes the
same value in both subsystems. This observation provides a strong constraint on
the eventual establishment of a granular equation of state. Moreover, among
various dynamical quantities studied, an effective diffusion coefficient is
shown to take the same value in the ``equilibrated'' subsystems, while the
structural relaxation times exhibit different behaviors. While a thermodynamic
system would lead to temperature equilibration, the present situation leads to a
decoupling of the mobility and the structural relaxation time, and only the
former seems to yield relevant information about the properties shared by
subsystems in steady states. 
Remarkably, this decoupling of the dynamics occurs in the vicinity of 
the {\it static} random loose packing.

\section{Experimental setup and protocol}

\begin{figure}
\centerline{\includegraphics[width=0.7\columnwidth]{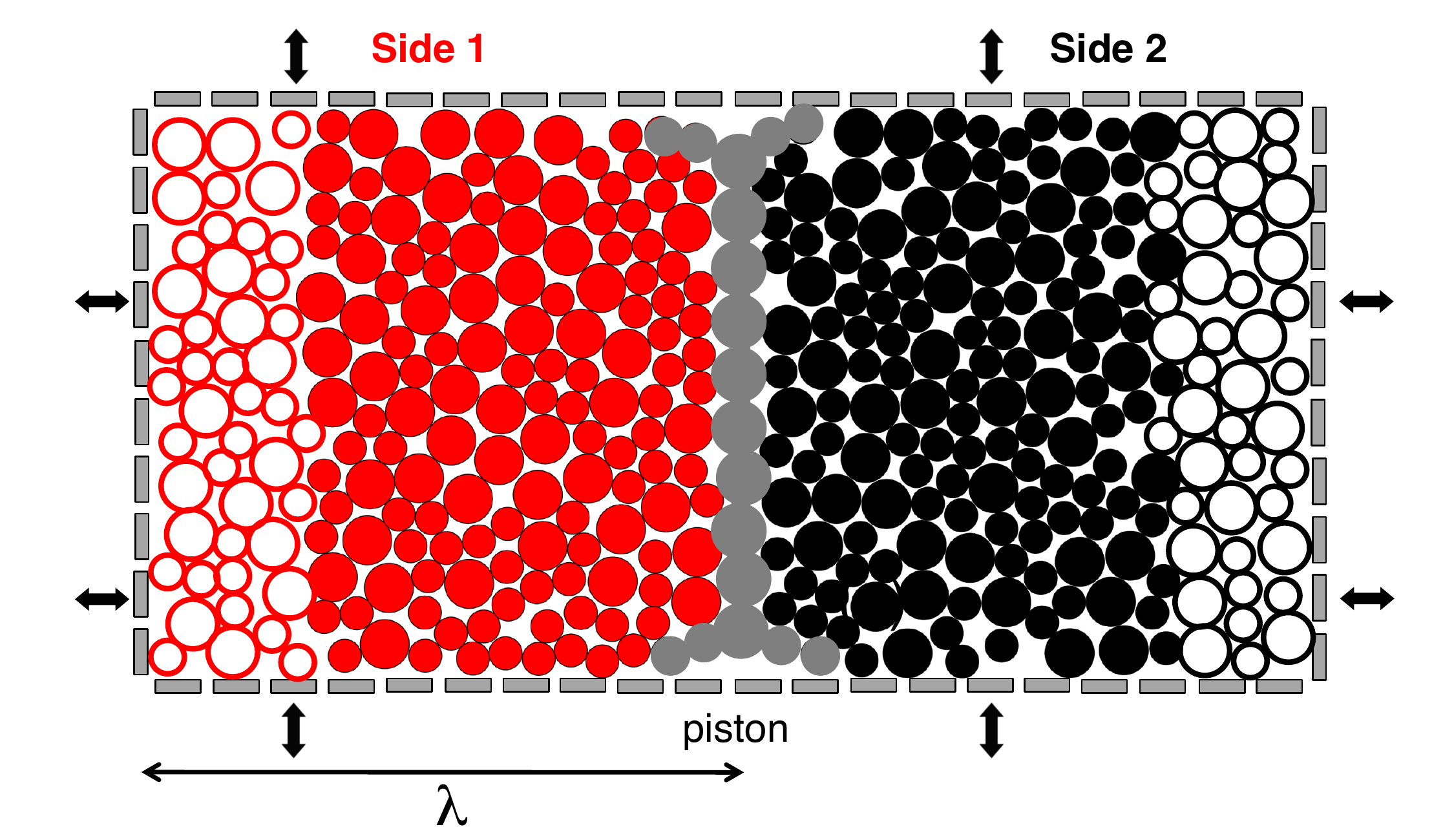}}
\caption{Apparatus schematic (1 m $\times$ 2 m), with particles drawn to scale.
Filled particles are taken from an experimental configuration and indicate the
size of the imaged region. $\lambda$ between $0$ and $1$ measures 
the fractional position of the piston.}
\label{fig:schematic}
\end{figure}

We study the equilibration of two adjacent dense bidisperse layers of disks on
an air table, as sketched in Fig.~\ref{fig:schematic}. A piston separates the
two
subsystems, and is constrained to move along the long axis of the air table by
protruding feet which impede rotation. On each side of the table, the system is
prepared with an identical number of particles $N$, with the ratio of small to
large particles fixed at $N_S = 2N_L$. This ratio suppresses crystallization and
provides roughly the same volume of large (diameter $d_L = 86$ mm) and small
particles ($d_S = 58$ mm). The particles within each subsystem have a different
set of material properties: on Side 1, the particles have restitution
coefficient $\epsilon_1 = 0.51 \pm 0.07$ and friction coefficient $\mu_1 =
0.85$; on Side 2, the particles have $\epsilon_2 = 0.33 \pm 0.03$ and $\mu_2 =
0.5$. Restitution coefficients were measured from isolated binary collisions;
friction coefficients are nominal values from the literature. Both sides utilize
standard plastic Petri dishes as the particles, with the difference in particle
properties achieved by encircling the particles on Side 1 with a rubber band. 
Because the sides of the particles slope inwards, the thickness of the rubber
band does not significantly change the radius of the particles; the mass of the
particles on Side 1 is increased by 7\%. 

The aggregate rearranges via an array of sixty electromagnetic bumpers which
form the walls of the system. These bumpers are triggered pairwise: bumpers
facing each other in the system fire at the same time in order to prevent net
momentum and torque injection. 
These pairs are triggered randomly via a pre-generated random sequence. 
Four pairs of bumpers are randomly fired every $0.1$ second, and travel 1 cm
into the granular pack;
the total time during which the bumpers stay in their forward position is
approximately 0.1 sec.

To quantify the long-time mobility of the particles, we take images at a
frequency which is low compared to the energy injection timescale, making usual
tracking techniques inoperative. Therefore, we have developed a tracking method
which identifies each particle by a unique tag. Each particle is marked with a
$3 \times 3$ array of colored dots which encodes two copies of a 4-bit, 4-digit
identifier, plus an error-correcting bit. The particles are located by their
circular rims and their identities are established using the tags, allowing
their positions
adjacent image frames to be connected into trajectories. We monitor the
positions of the piston and the inner 75\% of the particles with a CCD camera
mounted above the apparatus; we obtain a minimum of $10^4$ configurations for
each experiment. 

To understand the equilibration of the system on its approach to jamming, we
perform experiments at increasing values of $N$ while holding all other
variables constant. We adjust $N$ in increments of 2 small
and 1 large 
particle on each side, so that the two sides always have the same $N$. The
average 
packing fraction $\bar \phi$ calculated for the entire system is given by
\begin{equation}
\bar{\phi}\equiv
2\left(\frac{1}{\langle \phi_1 \rangle} + \frac{1}{\langle \phi_2
\rangle}\right)^{-1}
\end{equation}
as well as by the ratio of the total area of $N$ disks to the
total area of the air table. As we adjust $N$ from $183$ to $204$, this
corresponds to
 $\bar \phi = 0.768$ to $0.818$.

\section{Macroscopic observables}

\begin{figure}
\centerline{\includegraphics[width=0.6\columnwidth]{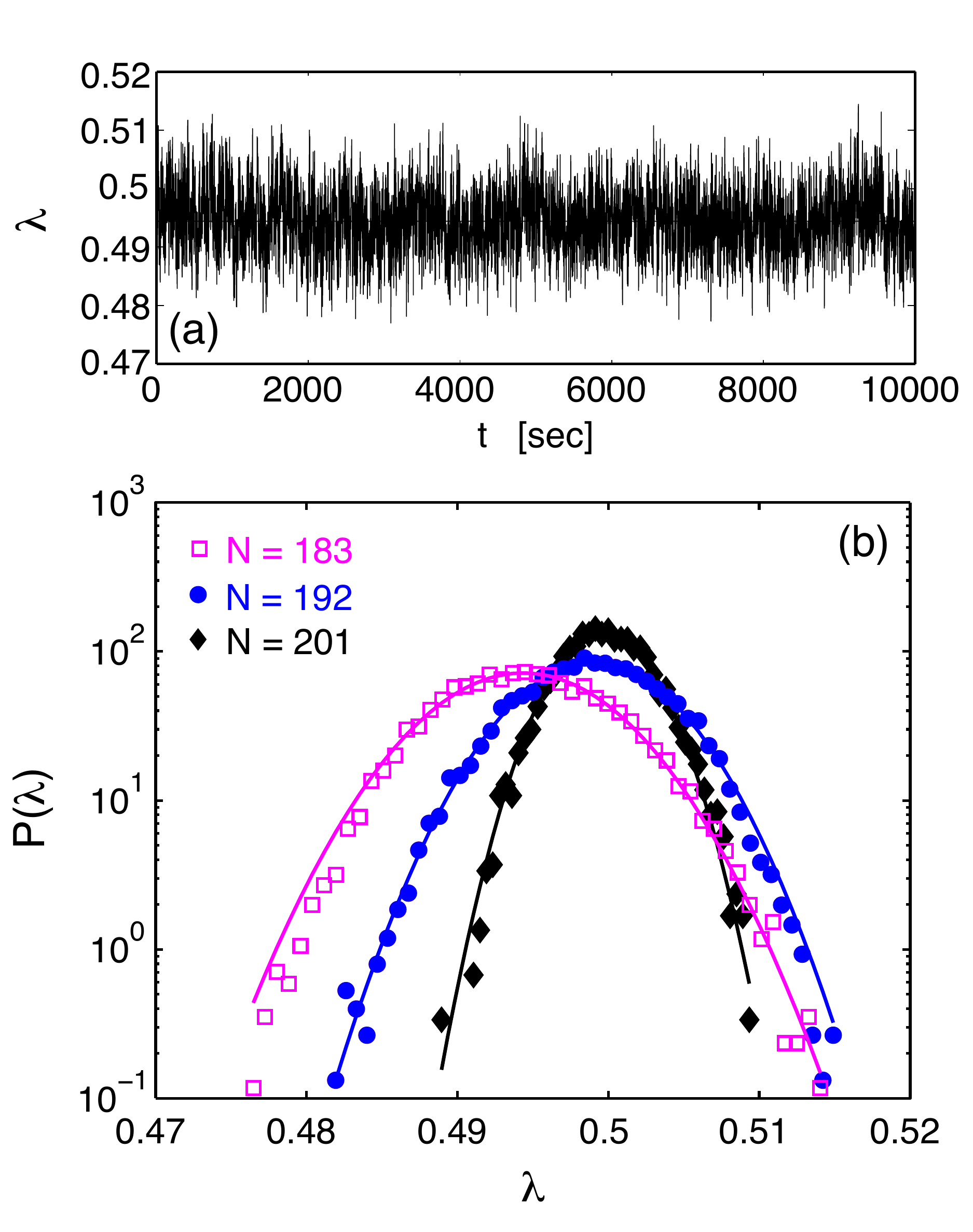}}
\caption{(a) Scaled piston position $\lambda(t)$ for $N=201$ ($\bar{\phi} =
0.812$). (b) Sample probability density functions of $\lambda$ (and Gaussian
fits) for $N=183$, $192$ and $201$ ($\bar{\phi}$ = 0.766, 0.789, 0.812). 
}
\label{fig:lambdaN}
\end{figure}

First we focus on the macroscopic  state of the system, namely
the volume occupied by each of the subsystems. Since the overall volume of the
system is conserved, we only need to consider the position of the piston scaled
by the length of the cell $0 < \lambda < 1$, where the origin of the axis is
taken on the side occupied by the particles that have been circled by a rubber
band, referred to as Side 1 in the following discussion. 

We first consider the time signal $\lambda(t)$ for a given value of $N$, an
example of which is displayed in Fig.~\ref{fig:lambdaN}a. After a transient time
$\tau_R$, the signal becomes stationary and
the system has reached steady state. This state is observed for all reported
values of $N$.  Note that $\tau_R$ is comparable to the
$\alpha$-relaxation time $\tau_{\alpha}$ of the assembly, i.e. the time it takes
for the
particles to diffuse further that their size \citep{Lois-2009-PMS}.
We hence consider the probability distribution of $\lambda$ in this
stationary regime and we find Gaussian statistics even for the largest $N$ 
(densest $\bar \phi$);
sample distributions are shown in Fig.~\ref{fig:lambdaN}b.

\begin{figure}

\centerline{\includegraphics[width=0.8\columnwidth]{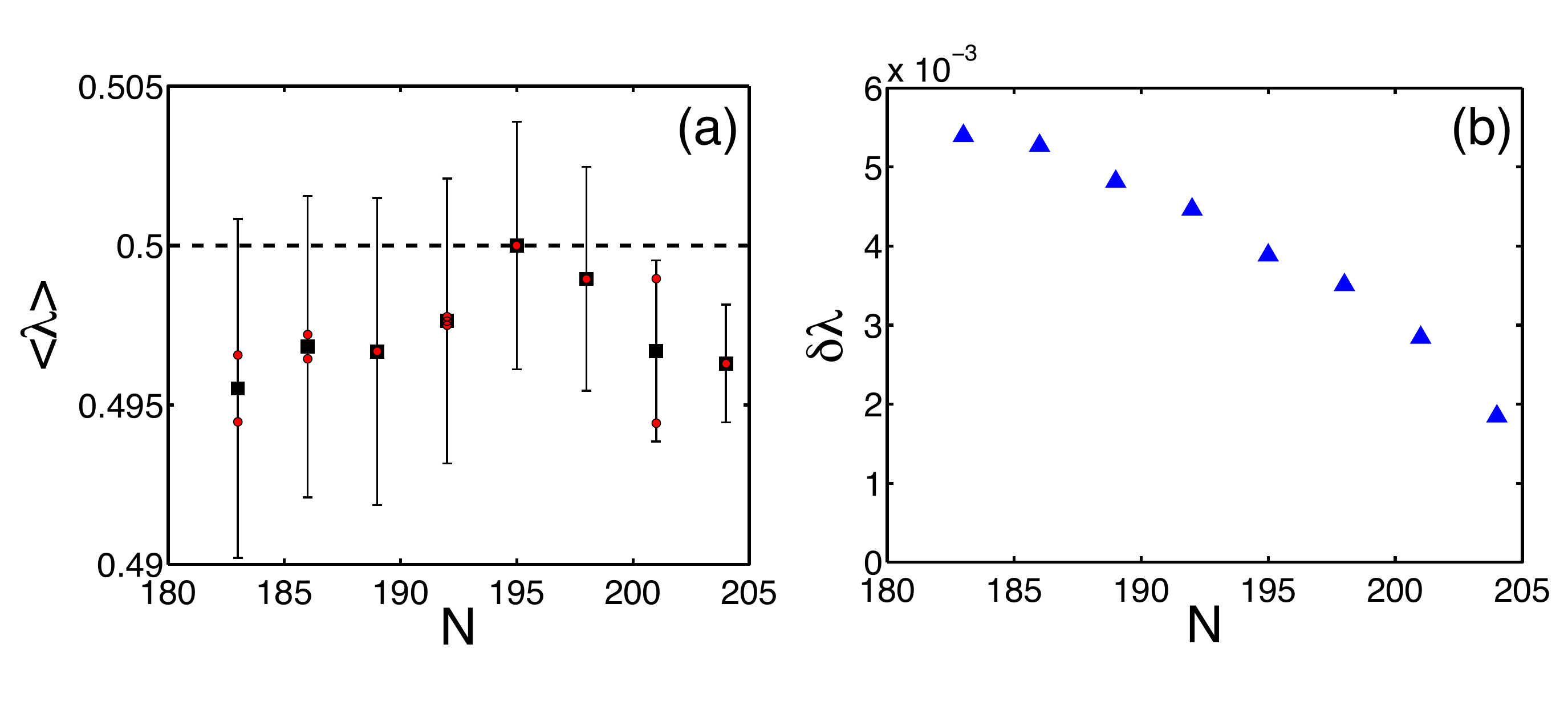}}
\caption{(a) $\langle \lambda \rangle$ (Side 1) as a function of $N$, with bars
representing the standard deviation. For some values of $N$, the experiment has
been repeated several times starting from different initial configurations, in
order to evaluate the errors; these measurements are shown as smaller symbols.
(b) Standard deviation $\delta \lambda$ as a function of $N$.
}
\label{fig:lambdaMoments}
\end{figure}

 These statistics are thus characterized by the first two moments.
We observe that the average piston position 
$\langle\lambda\rangle\equiv\langle \lambda\left(t\right)
\rangle_t$ is observed to be $\leq \frac{1}{2}$ for all values of $N$:  Side 1,
containing the particles with rubber bands, occupies less volume than the side
without.  We conclude that this systematic deviation from equal volumes
originates from the difference in material properties of the particles.
Furthermore, and rather surprisingly, $\langle\lambda\rangle$ is not a monotonic
function of $N$: it increases until it reaches $\frac{1}{2}$ for $N^*=195$ and
then
decreases for larger values of $N$. At this crossover, the packing fraction
is equal on both sides, $\phi^*=0.798$. The behavior of the system is
qualitatively different on each side of this transition point. For values of
$N\leq N^*$, the particles interact mostly through short binary collisions
 and the system looks
liquid-like. For $N\geq N^*$, the dynamics is dominated by multi-body
interactions and becomes much slower. In watching the system, intermittent force
chains \citep{Ferguson-2007-SHD} are be observed to carry the motion imposed
by
the boundaries into the bulk of the system, and some particles stay in contact
with one another on time scales larger than the typical collision time. However,
no permanent stress seems to be trapped in the system until it jams: this 
occurs at the highest explored value of $N$ as will be quantified below.

Interestingly, this transition at $\phi^*=0.798$ is close to, but perhaps
slightly
below, independent measurements of random loose packing for these particles:
$\phi^{RLP}_1 = 0.807 \pm 0.010$ and $\phi^{RLP}_2 = 0.812 \pm 0.006$. To
determine $\phi^{RLP}$ for each of the two types of particles, we placed the
table at a $0.3^\circ$ angle, and rained down single particles from the high end
to the low end to create the loosest packing accessible to us. These values are
upper bounds for random loose packing, for our protocol does not guarantee that
we reach the loosest possible stable packs.

The standard deviation of these distributions, $\delta\lambda \equiv
\sqrt{\langle\lambda^2\rangle-\langle\lambda\rangle^2}$, exhibits a simpler
behavior: it is a decreasing function of $N$, as can be seen in
Fig.~\ref{fig:lambdaMoments}b. This is to be expected from the fact that as $N$
increases
a smaller amount of free volume is available for the motion of the
piston. Notably, this quantity does not exhibit any special feature in the
vicinity of $N^*$. Finally, the magnitude of $\delta \lambda$ at the highest
explored value of $N$ closely matches the volume fluctuations imposed by the
bumpers. This mean that this macroscopic quantity becomes trivial when the
system approaches its jammed state, merely reproducing the microscopic
fluctuations imposed by the injection mechanism. 
However, even in the densest state explored the system is not strictly rigid:
particles continue to rearrange due to the impacts from the boundaries.

\section{Microscopic observables}

\subsection{Statics}

\begin{figure}
\centerline{\includegraphics[width=0.6\columnwidth]{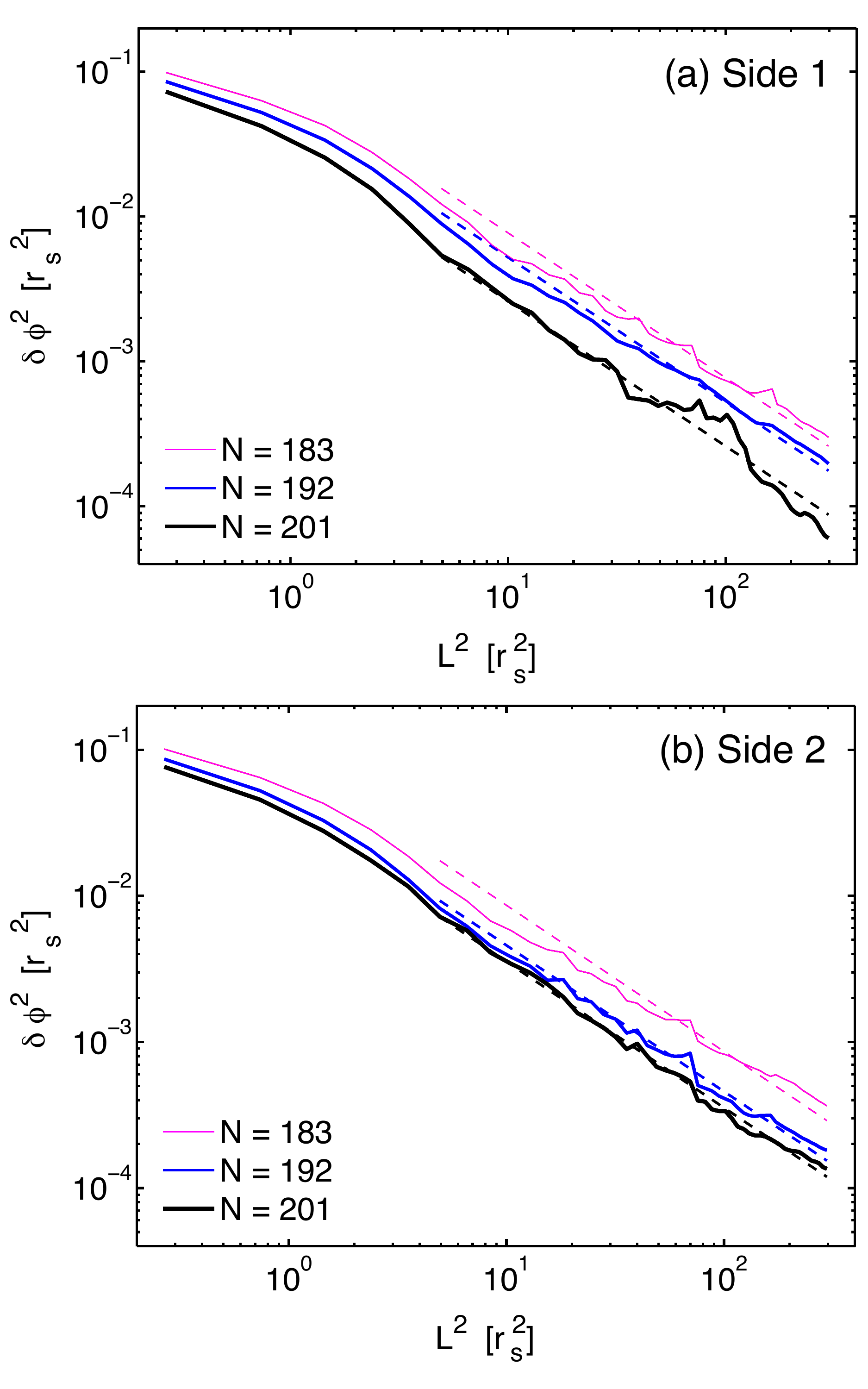}}
\caption{Variance $\langle \delta \phi^2 \rangle$ of the packing fraction
measured within squares of size $L$ for three values of $N$; dashed lines are
$\langle\delta\phi^2\rangle_0 / L^2$.}
\label{fig:dphi}
\end{figure}

We now turn to the microscopic information provided by particle-tracking. In
order to extract a well-defined, size-independent microscopic quantity from the
packing fraction field, we analyze the spatial and temporal fluctuations of the
average
local packing fraction $\phi$ over windows of increasing size, where $\phi$ is
defined as the ratio of the number of pixels occupied by the particles to the
total number of pixels in a given region. For each subsystem, we measure $\phi$
over boxes of size $L$ ranging from a few $r_s = \frac{1}{2} d_s$
 up to half the system size. For
$L \gtrsim 2r_s$, $\phi(L)$ converges to a constant. Moreover, the variance
$\langle \delta \phi^2 \rangle$ scales approximately as $L^{-2}$, as shown in
Fig.~\ref{fig:dphi}. Such scaling behavior is expected from the central limit
theorem provided there are no long-ranged spatial correlations in the packing
fraction
field. Since the packing fraction does not exhibit such correlations, it is
suitable for a thermodynamic-like analysis despite the relatively small number
of particles. Furthermore, we obtain an $L$-independent measure of the
variance of $\phi$ by averaging $\langle \delta\phi^2 \rangle _0 \equiv \langle
\delta \phi^2 \rangle  L^2 $ over all $L > 2 r_S$.

\begin{figure}
\centerline{\includegraphics[width=0.8\columnwidth]{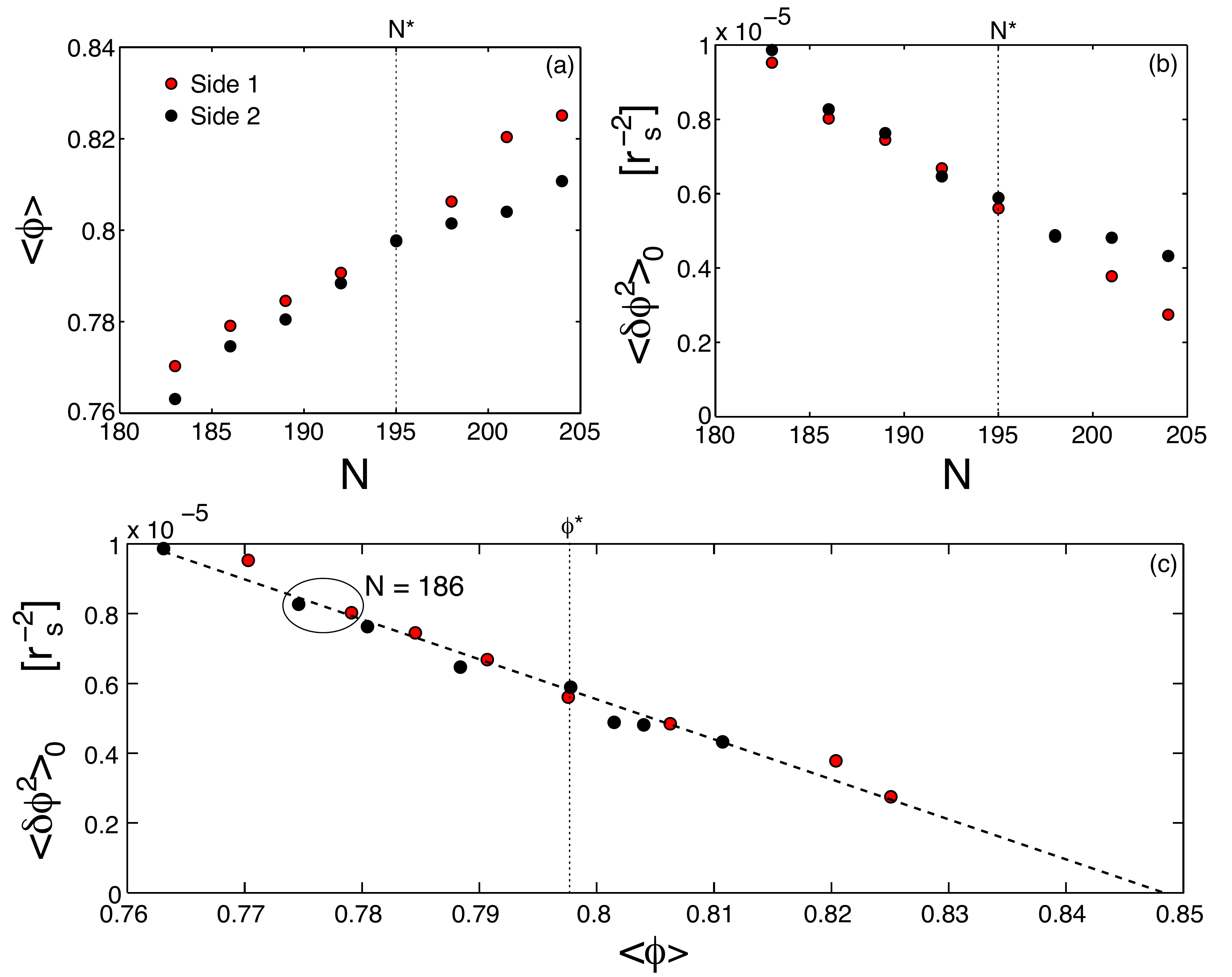}}
\caption{(a) Packing fraction $\langle \phi \rangle$ and (b) normalized packing
fraction fluctuations  $\langle \delta\phi^2 \rangle_0$ as a function of number
of particles on each side. (c) Data from parts (a) and (b) combined on a single
plot without regard for $N$. Ellipse encloses two points obtained from Side 1
(red) and Side 2 (black) for a single run with $N=186$. Dashed line is a linear
fit to data from both sides, showing an intercept at $\phi_J = 0.848$.
}
\label{fig:equi}
\end{figure}

These two packing fraction statistics are plotted as a function of $N$ in
Fig.~\ref{fig:equi}: $\langle \phi \rangle$ is higher for
Side 1, in agreement with our macroscopic measurement. Similarly, at $N=195$ the
packing fractions are equal ($\langle \phi \rangle = \bar{\phi} = 0.798$ for
both sides). Different values of the normalized $\phi$ fluctuations $\langle
\delta\phi^2\rangle_0$ are observed on the two equilibrated sides, indicating
that the statistics of this quantity depend on the material properties.
Moreover, the fluctuations on each side are found to be decreasing functions of
$N$, as shown in Fig.~\ref{fig:equi}b. As for the fluctuations of $\lambda$,
this can be understood by the fact that as $N$ increases, the amount of free
volume to be distributed among the particles decreases. Finally, the discrepancy
in the packing fraction statistics between the two sides gets larger on approach
to jamming at large $\bar \phi$ (large $N$), especially as the number of
particles is increased above $N^*$.

To examine the relevance of these two state variables, we plot the dependence of
$\langle \delta\phi^2\rangle_0$ on $\langle \phi \rangle$,  parametrized by $N$
(see
Fig.~\ref{fig:equi}c). Remarkably, the data from both sides
fall onto a single master curve: the fluctuations in local $\phi$ are
 {\itshape insensitive} to the
material properties of the particles. Within the explored range of $N$, this
master curve is approximately linear and the extrapolation would intersect
$\langle\delta\phi^2\rangle_0=0$ at $\langle \phi \rangle = \phi_J = 0.845$,
which is compatible with the rigidity transition reported in other bidimensional
systems \citep{OHern-2003-JZT, Majmudar-2007-JTG, Lechenault-2008-CSH}. This
result therefore stands as a starting point for a granular equation of state.
It is notable that, as in Fig.~\ref{fig:equi}b, 
the transition at $N^* = 195$ does not appear as a 
feature.

\subsection{Dynamics}

\begin{figure}
\centerline{\includegraphics[width=0.7\columnwidth]{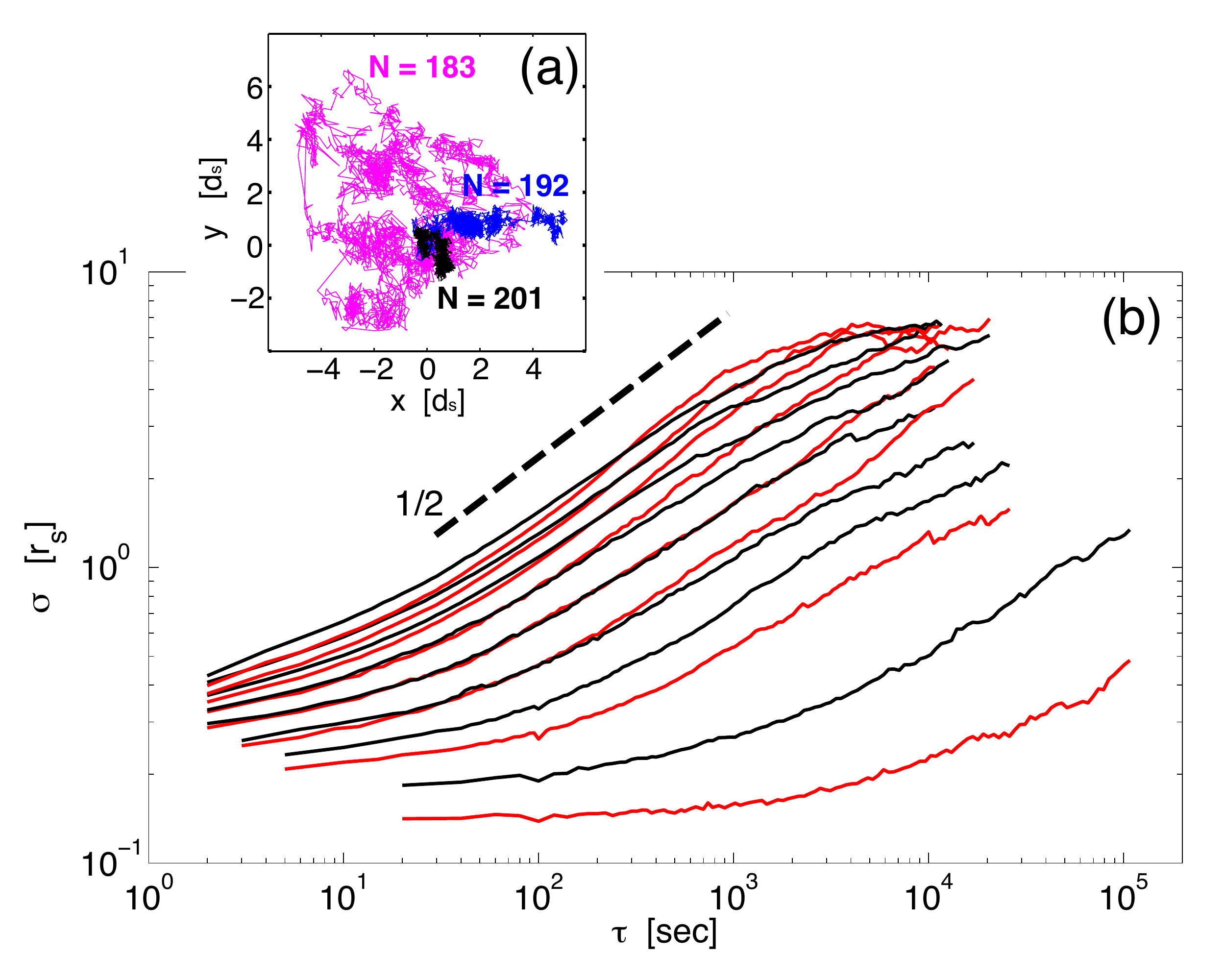}}
\caption{(a) Trajectories of a single particle (from Side 1) for a duration of
$10^4$ sec. (b) Diffusion length $\sigma$ as a function of lag-time $\tau$. Red
lines are for Side 1; black lines are for Side 2. Dashed line is $\sigma \propto
\tau^{\frac{1}{2}}$ to guide the eye.
}
\label{fig:diffusion}
\end{figure}

We now turn to the microscopic characterization of the dynamics in our system
as the system transitions from a
liquid-like to a solid-like state. To monitor this transition, we use
particle-trajectories to compute the average diffusion distance as a function of
lag-time $\tau$, defined as
\begin{equation}
\sigma(\tau) \equiv \sqrt{\langle\|\vec{r}_i\left(t+\tau\right) -
\vec{r}_i\left(t\right)\|^2\rangle_{i,t}}
\label{e:sigmadef}
\end{equation}
where $\vec{r}_i\left(t\right)$ is the vector position of particle $i$ at time
$t$. Averages are computed over the ensemble of all particle trajectories at all
times.

Fig.~\ref{fig:diffusion} shows how the r.m.s.
displacement $\sigma (\tau)$ and the corresponding particle trajectories
vary as increasing $N$ brings the system towards
jamming. At $N=183$ (low $\bar \phi$) a particle explores a region several $r_s$
wide, and $\sigma$ saturates at long time scales due to finite system size. On
the other hand, at $N=201$ (high $\bar \phi$), $\sigma$ exhibits  sub-diffusive
behavior
at short $\tau$, and caging effects are significant. As the dynamics of
the system slow down at large $N$, we increase the duration of the experiment
from twenty hours up to fifty-five hours for the densest packing in order to
sample a significant set of configurations. However, even at this extended
acquisition time, the diffusion length at $N=201$ barely passes $1 r_s$. From
the glass transition point of view, the system is completely jammed. 

\begin{figure}
\centerline{\includegraphics[width=0.8\columnwidth]{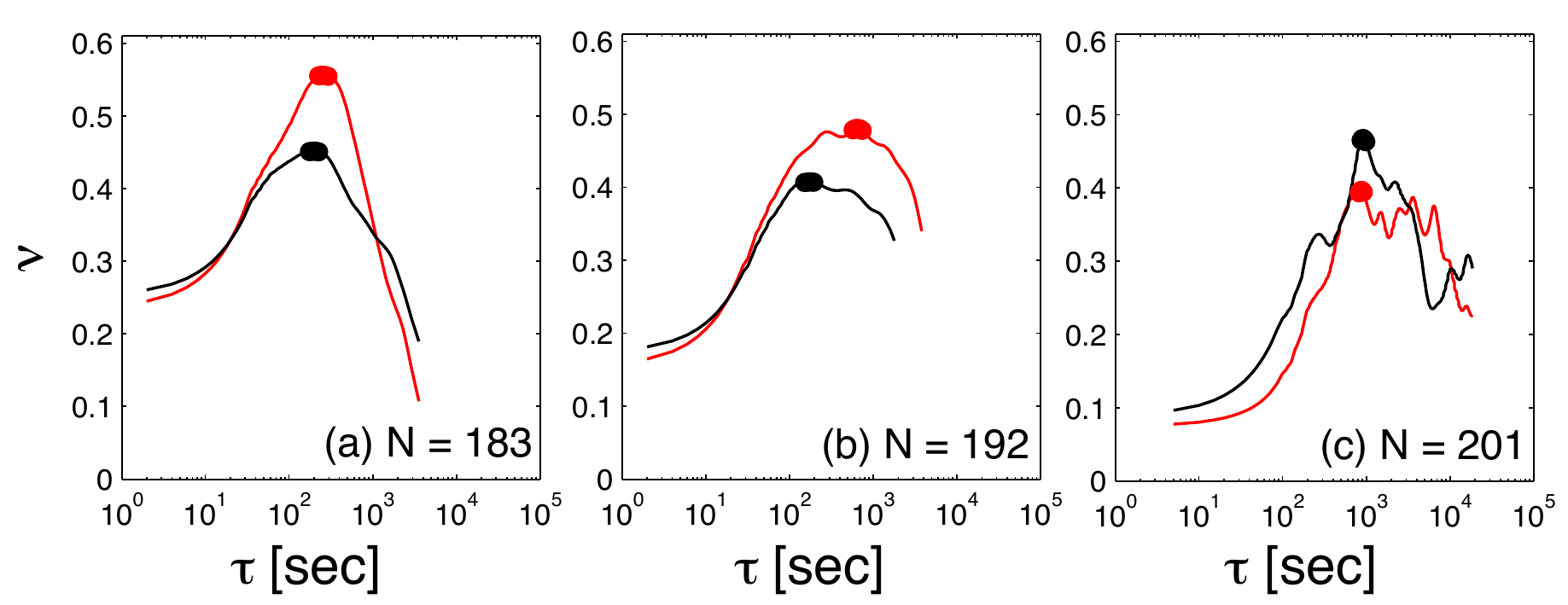}}
\caption{Local exponent $\nu$ as a function of lag-time $\tau$ for  $N=183$, 
$192$ and $201$. The thick region indicates the averaged maximum used to obtain 
a diffusion constant $D$ from Eq.~\ref{e:nu}.}
\label{fig:nu}
\end{figure}

\begin{figure}
\centerline{\includegraphics[width=0.8\columnwidth]{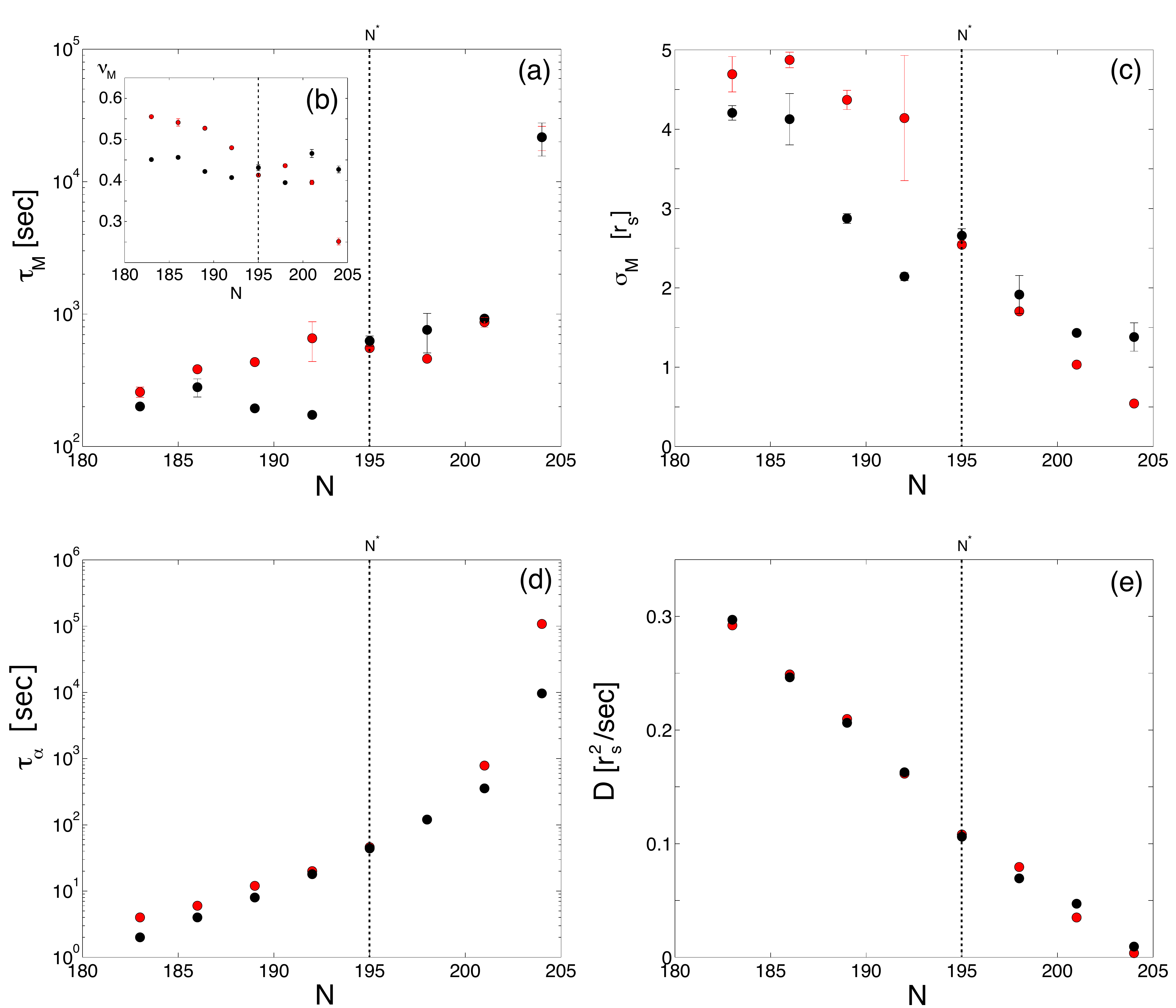}}
\caption{Maximum exponent $\nu_M$, and corresponding values of $\sigma_M$ and
$\tau_M$ as a function N for the two grain species. (c) Diffusion coefficient
$D$ and (d) structural relaxation time $\tau_{\alpha}$ as a function of $N$. 
}
\label{fig:diffusionN}
\end{figure}

In order to extract meaningful information from $\sigma(\tau)$, including
establishing
whether a truly diffusive regime is ever reached, we
compute the local slope 
\begin{equation}
\nu \equiv \frac{\partial (\log \sigma) }{\partial (\log \tau)}.
\label{e:nu}
\end{equation}
over a range of $\tau$ that allows for a reasonable estimation of the local
slope.
Interestingly, for all studied values of $N$ this local exponent exhibits a 
maximum, as illustrated in Fig.~\ref{fig:nu}. This maximum is consistently close
to $\frac{1}{2}$, as can be seen in Fig.~\ref{fig:diffusionN}b. However, 
the exponent $\nu$ is larger for Side 1 (rubber bands) than Side 2 (bare) 
for $N < N^*$. Above $N^*$, Side 1 changes from super-diffusive ($\nu >
\frac{1}{2}$)
to sub-diffusive ($\nu < \frac{1}{2}$) behavior. 
In contrast, Side 2 is sub-diffusive  for all values of $N$.

We have extracted the characteristic time and length scales $\tau_M$
and $\sigma_M$ at which this maximum is reached, plotted in
Fig.~\ref{fig:diffusionN}. For both sides of the system, the 
time scale at which maximally-diffusive behavior is reached rises by two 
orders of magnitude as 
$N$ approaches the jammed state, and the length scale falls.
These quantities indicate the occurrence of a dynamical
crossover at $N^*$ since
$\sigma_M$, $\tau_M$, and $\nu$ all display different behavior above and 
below this value. 
Side 1 is more mobile for $N < N^*$ and Side 2 is more mobile above.
Interestingly, this transition occurs when $\sigma_M$ is close to the mean 
diameter of the grains  ($\sigma_M = 2 r_s$). This suggests that
the local rearrangement mechanism governing the dynamics at the particle scale
changes near $N^*$. 

We use the behavior near $(\sigma_M, \tau_M)$ to extract quantities which 
quantify the diffusive behavior of the system. The structural relaxation time
$\tau_{\alpha}$ is obtained from the time at which
$\sigma\left(\tau_{\alpha}\right)$ crosses $r_s$,
meaning that the particle has diffused beyond its radius. 
A diffusion coefficient $D$ is estimated by solving 
\begin{equation}
\sigma_M = D \, \tau_M^{\frac{1}{2}} 
\label{e:diffusion}
\end{equation}
for $D$. Since the dynamics of the system are not properly diffusive, this
diffusion 
coefficient should be interpreted merely as a measure of the
displacement amplitude at the time of maximum mobility. Astonishingly, 
as can be seen in Fig.~\ref{fig:diffusionN},
this definition for $D$ provides a dynamical parameter that takes approximately
the {\itshape same} value in each subsystem regardless of $N$.
 
Both $D$ and  $\tau_\alpha$ indicate a slowing down for increasing $N$, as shown
in Fig.~\ref{fig:diffusionN}. On both sides, $D$ decreases monotonically to
nearly zero for our last data point, indicative of the system's cessation of
rearrangement (jamming). Meanwhile, as is usually observed in the vicinity of
the glass/jamming transition, the structural relaxation time for both species of
particles soars over multiple orders of magnitude. The plot of $\tau_\alpha(N)$
shows that both above and below $N^*$, Side 1 (rubber band particles) exhibit
longer relaxation times than Side 2. As for $\nu$, $\sigma_M$, and $\tau_M$,
the $N^*$ crossover point is special in that these quantities each take the 
same value on both sides. No dramatic changes take place for $D(N)$
near $N^*$: it is remarkably equal in each subsystem for all values of $N$.

\begin{figure}
\centerline{\includegraphics[width=0.8\columnwidth]{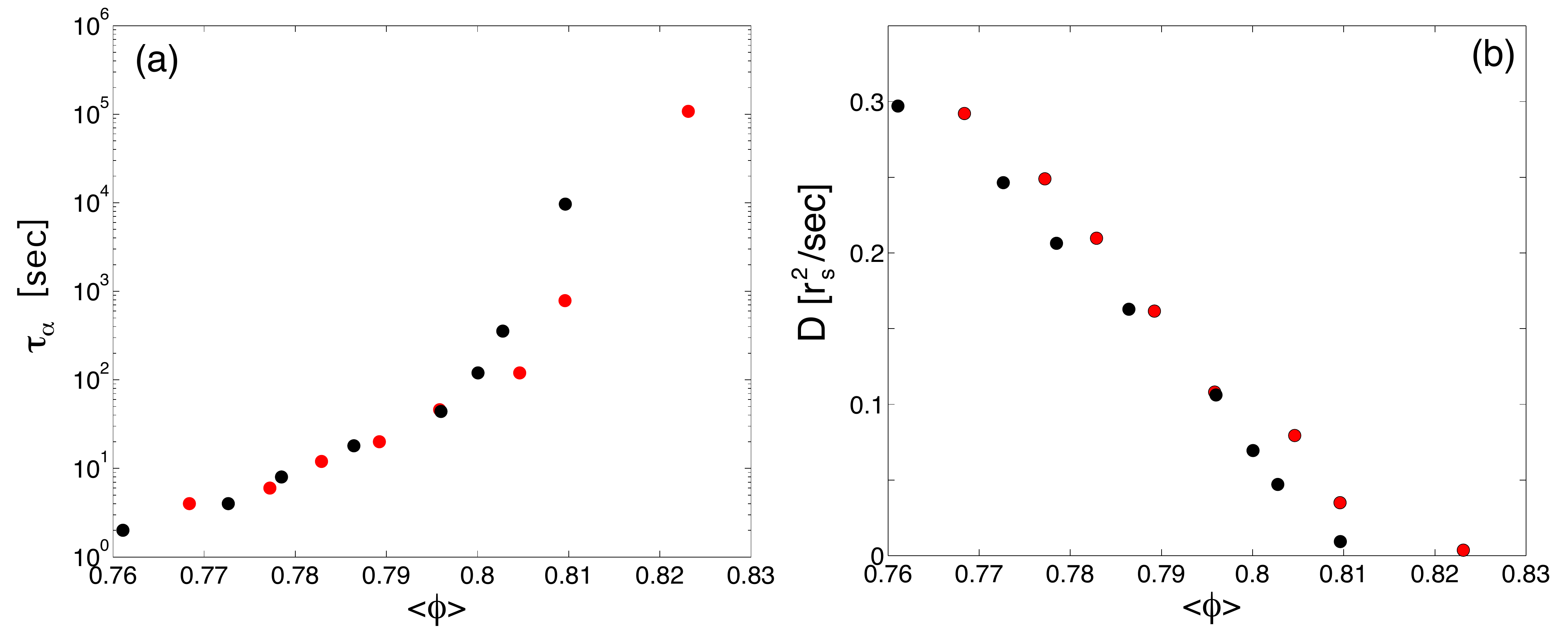}}
\caption{(a) Diffusion coefficient $D$ and (b) structural relaxation time
$\tau_{\alpha}$ as a function of $\langle \phi \rangle$ (without regard for
$N$).}
\label{fig:diffusionPhi}
\end{figure}

Since the timescale $\tau_\alpha$ is observed to be
sensitive to $N^*$, we investigate the extent to 
which the packing fraction is relevant to the dynamics. 
Therefore, we plot both $D$ and $\tau_\alpha$ as a function of $\langle \phi
\rangle$ in Fig.~\ref{fig:diffusionPhi}. As in Fig.~\ref{fig:equi}, pairs of
data points from the same run are plotted at their corresponding values of
$\langle \phi \rangle$. This parametrization predicts that $D(\phi)$ vanishes at
different values for the two different particle types, with $\phi_J^{(1)} <
\phi_J^{(2)}$. In examining  $\tau_\alpha(\phi)$ for the two sides, there is a
common (material-independent) branch of the curve for 
 $\langle \phi \rangle \lesssim \phi^*$
which increases gradually. The value of $\langle \phi \rangle$ at which
$\tau_\alpha$ leaves this branch differs for the two types of particles, with
Side 1 again having its transition at lower $\langle \phi \rangle$. This
transition is associated with local rearrangement mechanisms, and likely
corresponds to the packing fraction 
at which single-particle rearrangements are no longer possible
\citep{Aste-2005-VAD}.

\section{Discussion and Conclusions}

We have explored the stationary states of subsystems able to exchange volume,
following them from a liquid-like to a jammed state. In order to examine the
extent
to which state variables and the density of states are useful tools to
understand
this transition, we would need to understand the role of energy, entropy, and
our ability to identify the corresponding microstates. An analysis
based on the generation of entropy through the braiding of trajectories 
 \citep{Thiffeault-2005-MTC,Puckett-2009-GEM} establishes the
chaotic nature of the internal dynamics in our system. This, together with the
stationarity, ensures the existence of an invariant measure, namely the SRB
measure \citep{Eckman-85}, from which our temporal averages are guaranteed to
correspond to ensemble averages. This provides a strong basis for the
statistical
analysis provided above and lends a thermodynamic-like status to the quantities
extracted from our equilibration experiments. Within such a framework, our
results 
suggest further interpretation.

In Fig.~\ref{fig:equi}, we have uncovered an unexpected one-to-one relationship
between the average local packing fraction $\langle \phi \rangle$ and its
fluctuations $\langle \delta\phi^2 \rangle_0$. This striking
relationship, in which both $N$ and the material properties collapse on
a single curve, should be a prediction from an as-yet unknown equation of state.
Interestingly, the function  does not appear to be sensitive to the transition
at $\phi^* \approx \phi^{RLP}$ which is a prominent feature in the equilibration
measurements of the piston position (see Fig.~\ref{fig:lambdaN}).

To make use of the relationship, it is tempting to define a variable 
\begin{equation}
Q \equiv \frac{\langle \delta \phi^2 \rangle_0}{\langle\phi\rangle - \phi_J}
\end{equation}
which is intensive, $\phi$-independent, and takes the same value in both
subsystems for all $N$. However, the value of $Q$, which we conjecture to be
related to the energy injection and/or dissipation rates in the system, is not
sufficient to specify the equilibrated volumes. This suggests that another
parameter
is needed to fully characterize the state of a dynamically evolving dense
granular pack, much as both temperature and pressure equilibration are needed to
solve the equivalent problem within classical thermodynamics. The present
experiment does not allow us to measure the instantaneous kinetic and potential
energies, but future studies of particle kinetics would allow further
investigation of such parameters.

Diffusion measurements provide an alternative framework in which to understand
the
behavior of our system and hence the equilibration process. We have observed
that the piston
position $\lambda$, relaxation time $\tau_\alpha$, and 
the maximum mobility length and timescales
$\left(\sigma_M,\tau_M\right)$ are sensitive to the transition at $\phi^*$ (see
Fig.~\ref{fig:lambdaN}) which is close to the static $\phi^{RLP}$. These changes
in the dynamics suggest that the transition corresponds to two qualitatively
different dynamical regimes. 
For $\phi < \phi^*$, the particles primarily interact through short binary
collisions
and there is weak coupling between the particles. Because the system is
collision-dominated, it is likely that the restitution coefficients of the
particles are an important consideration in the dynamics. If so, it would
explain the observation that for Side 1 (larger coefficient of restitution),
$\lambda$ increases with $N$. 
For $\phi > \phi^*$, long multibody interactions dominate the dynamics, and we
visually observe signs of intermittent force chains \citep{Ferguson-2007-SHD}
which strongly couple both particle-particle and particle-bumper interactions.
It
is likely that the higher friction coefficient on Side 1 significantly increases
the structural relaxation time of the corresponding assembly, and the dynamics
are more sensitive to the frictional properties of the particles. 
Interestingly, all measured quantities are approximately equal in each subsystem
at $N^*$.

As the system approaches $\phi_J$, the persistence time of the contact network
increases, and the internal response of the particles is more correlated. For
the largest studied value of $N$, the packing fraction and overall volume
fluctuations are are imposed by the boundaries. Hardly any rearrangement occurs
on experimentally-accessible time scales, but extended internal vibrations
can be observed: the behavior of the system is thus comparable to that of a
solid.  

From a thermodynamic point of view, it is remarkable that the effective
diffusion
coefficients defined above are the only measured quantities to take the same
value in the equilibrated subsystems, whereas the relaxation times
$\tau_{\alpha}$ become very different from each other beyond $N^*$. 
As a matter of fact, in
equilibrium systems, the Stokes-Einstein relation states that the product of the
diffusion
coefficient and the viscosity (or, equivalently, the relaxation time),
is proportional to temperature \citep{Kubo-1966-FDT}. Hence, only the product 
$D\tau_\alpha$, 
and not $D$ itself,
is expected to equilibrate. Here, we find that there is a decoupling between the
relaxation time and the diffusion coefficient beyond $N^*$, which is reminiscent
of the decoupling reported in the vicinity of the glass transition in
supercooled liquids \citep{fujara1992translational,debenedetti2001supercooled},
the latter equilibrating
while the former emphasizes the difference in material properties between the
two sides. This phenomenon is yet another common feature between dense granular
materials and supercooled liquids, and seems to indicate that
fluctuation/response based temperatures might be relevant below
$\phi^{RLP}$.\cite{Liu-1998-NDJ}

Thus, we observe that both packing fraction (or free volume) and material
properties play a key role in determining the state of the system. In static
granular 
systems, the Edwards ensemble \citep{Edwards-1989-TP} provides a framework for
understanding how volume 
plays a central role in describing the density of states and associated
Boltzmann-like thermodynamic properties. While our system is dynamic, it is
interesting to ask to what extent the equilibration of volume between two
subsystems exhibits features reminiscent of the Edwards formalism. 
The overall volume of our system is a conserved quantity, the dynamics explore a
set of configurations (although not strictly mechanically-stable), and we
achieve stationary distributions of system properties such as
$\langle \phi \rangle$. Were volume the dominant state variable setting the
density of states for this system, we would
have observed $\lambda = \frac{1}{2}$, independent of $N$. Instead, the material
properties of the particles play an important role in dynamically determining
the state of the system, and we observe that in general $\lambda \ne
\frac{1}{2}$.

\bigskip

In this paper, we have experimentally  studied  the steady state statistics of
two
model granular subsystems differing in their material properties put in contact
through a mobile wall. We have uncovered several features that 
constrain any specification of a granular equation of state. 
First, most static
and dynamical quantities are affected by a dynamical crossover which occurs at a
packing fraction close to the (static) random loose packing. We interpret
this crossover as separating the weak and
strong coupling regimes in the behavior of the system. However, both the
macroscopic volume
fluctuations and the diffusion coefficients are insensitive to this transition.
Second, we have found
that the packing fraction fluctuations are in a one-to-one
correspondence with the average packing fraction, independent of the
material properties of the grains. 
These fluctuations drop to zero on the approach to jamming.
Finally, the diffusion properties of the particles reveal a glassy feature known
as decoupling which arises above random
loose packing. However, in spite of the fact that the structural relaxation
timescale 
becomes both large and material-dependent for dense systems, the effective
diffusion 
coefficient nonetheless takes the same value in each subsystem. This suggests
that this coefficient is an important intensive parameter describing the state
of the system.
A theoretical understanding of these last two findings could come from 
the shrinking density of states on the approach to jamming. It is intriguing
that the packing 
fraction fluctuations (geometry) suggest material-independence while 
the relaxation timescales (dynamics) suggest material-dependence. 

\newpage

\providecommand*{\mcitethebibliography}{\thebibliography}
\csname @ifundefined\endcsname{endmcitethebibliography}
{\let\endmcitethebibliography\endthebibliography}{}

\end{document}